\documentclass[12pt]{article}
\usepackage{amsmath,latexsym,amssymb,epsfig}
\newcommand{\beeq}{\begin{equation}}
\newcommand{\ene}{\end{equation}}
\newcommand{\bea}{\begin{eqnarray}}
\newcommand{\ena}{\end{eqnarray}}
\newcommand{\no}{\noindent}
\newcommand{\nb}{\nonumber}

\newcommand{\Bigbreak}{\par \ifdim\lastskip < \bigskipamount \removelastskip
\fi \penalty-300 \vskip 10mm plus 5mm minus 2mm}

\newcommand{\de}{\partial}

\newcommand{\e}{{\rm e}}

\newcommand{\Si}{{\cal S}}
\newcommand{\M}{{\cal M}}

\newcommand{\B}{{\cal B}}

\newcommand{\ha}{\frac{1}{2}}

\newcommand{\bM}{{\bf M}}
\newcommand{\R}{\mathbb{R}}
\newcommand{\bx}{{\bf x}}
\newcommand{\bp}{{\bf p}}

\newcommand{\der}{\partial }

\begin{document}

\thispagestyle{empty}

\rightline {SNS-PH/00-11}
\rightline {IFUP-TH 21/2000}

\vskip 0.5 truecm
\centerline {\Large \bf Localization of Quantum Fields on Branes}
\vskip 1 truecm
\centerline {\large \rm Mihail Mintchev}
\medskip
\centerline {\it Istituto Nazionale di Fisica Nucleare, Sezione di Pisa}
\centerline {\it Dipartimento di Fisica dell'Universit\`a di Pisa,}
\centerline {\it Via Buonarroti 2, 56127 Pisa, Italy}
\bigskip
\medskip
\centerline {\large \rm Luigi Pilo}
\medskip
\centerline {\it Scuola Normale Superiore,}
\centerline {\it Piazza dei Cavalieri 7, 56126 Pisa, Italy}
\centerline {\it Istituto Nazionale di Fisica Nucleare, Sezione di Pisa}
\medskip
\vskip 1.5 truecm
\centerline {\large \it Abstract}
\medskip
A mechanism for localization of quantum fields on a $s$-brane, representing
the boundary of a
$s+2$ dimensional bulk space, is investigated. Minkowski and AdS bulk
spaces are analyzed.
Besides the background geometry, the relevant parameters controlling the
theory
are the mass $M$ and a real parameter $\eta$,
specifying the boundary condition on the brane. The importance of exploring
the whole
range of allowed values for these parameters is emphasized. Stability in
Minkowski space requires
$\eta \geq -M$, whereas in the AdS background all real $\eta$ are permitted.
Both in the flat and in AdS case, the induced field on the brane is a
non-canonical
generalized free field. For a suitable choice of boundary condition,
corresponding to the
presence of a boundary state, the induced field on the brane mimics standard $s+1$ dimensional physics. In a certain range of $\eta$, the spectral function i
n the the AdS case is
dominated by a massive excitation, which imitates the presence of massive
particle
on the brane. We show that the quantum field induced on
the brane is stable.




\bigskip
\bigskip

\centerline {June 2000}
\vfill \eject

\section{\bf Introduction}
The scenario of a physical space-time with large extra dimensions is now
attracting much interest.
According to this proposal \cite{ex}, our world is confined in a
four-dimensional defect, a 3-brane,
embedded in a higher dimensional space $\M$, called bulk space.
The extra dimensions have sub-millimeter size, which can be tested, in
principle,
in future collider experiments \cite{Ricc}. From a geometric point of view, the
bulk space is a product
of the Minkowski space $\mathbf{M}_{3+1}$ with a suitable compact space
like in the  Kaluza-Klein (KK)
program. Further insights come from the Randall-Sundrum solution
\cite{RS1,RS2}, where ${\cal M}$ is a slice of the anti-de Sitter
(AdS) space-time in five dimensions. Localization of gravity is realized
through the ``zero mode'' of the five-dimensional
gravitational field, which is appreciably different from zero only close to
the 3-brane of our world. Despite that the extra fifth dimension is
infinite, the localization mechanism
works also in the non-compact case \cite{RS2} and Newton's law is
essentially recovered \cite{garriga}.
Another class of models \cite{grs,ross}, potentially interesting for the
cosmological
constant problem, is obtained when
the bulk space has an additional internal negative-tension brane, which
separates an AdS region from
a flat region. In this case, the Newtonian potential on the world brane is
recovered only at
intermediate scales \cite{grs,csaki}, however there are some issues
\cite{porr,grs1,csaki1,me} related with
unitarity and stability which are still not completely understood.

In this paper we will be concerned with the mechanism responsible for the
localization of a quantum scalar field
on a brane, representing the boundary of an infinite bulk space. More
precisely, we address the following quantum field theory problem.
Let $\Phi $ be a local quantum field with prescribed dynamics, propagating
in a bulk space $\M$ with a nontrivial boundary $\der\M$.
The problem is to construct and investigate the field
$\varphi $
induced by $\Phi$ on $\der \M$. Referring in what follows to $\varphi$ as
the induced field,
our main goal is to study the general mechanism allowing to localize a
quantum field on $\der \M$.
We also analyze the interplay between locality on $\M$ and $\der \M$
and the relationship between the mass spectra of $\Phi$ and $\varphi$.
Our interest in this sort of quantum boundary value problem is
not new \cite{LMZ,LM,GLM}. In order to establish a contact with the
previous work, we start by considering a flat bulk space which is both
instructive and provides a reference model for the study of the more
involved AdS background.
Our investigation shows that the behavior of $\varphi $
is controlled essentially by two parameters: the bulk mass $M\geq 0$ and a
real
parameter $\eta $ (with dimension of mass), which specifies a generic mixed
boundary condition on $\der \M$.
One of the main differences between the flat and the AdS bulk space
concerns the allowed values of $\eta$. In the AdS case, $\eta$ can take any
real value, whereas for flat $\M$ the requirement of stability
implies $\eta \geq - M$. It turns out that
besides the continuum KK modes, for a suitable
value $\eta_b(M)$ of $\eta $, the spectrum contains a ``zero mode'',
decaying
exponentially out of the brane and generating massless excitations on it.
This feature, which was already established in the case $M=0$, extends
therefore
also to $M>0$. A remarkable consequence of this fact is that short range
interactions in the flat bulk space $(M>0)$ induce long range
interactions on the brane when $\eta = \eta_b$.
For $M>0$ and $\eta \not= \eta_b$ the
induced field $\varphi$ has a mass-gap in flat bulk space and no gap
in the AdS background. The potential between two static sources on the
brane behaves accordingly.

We stress that for establishing these results it
is fundamental to have under control the whole range of the parameter
$\eta $.
Some partial results for $\eta = 0$ have been derived recently
in \cite {rub1,lisa}. Lifting this restriction is however essential, because it
turns out for instance that $\eta_b \not= 0$ for $M>0$ .
The study of the problem for general admissible $\eta $ is therefore crucial
and is among the main achievements of this paper. Another interesting aspect,
discussed below, is the behavior of the continuum mass spectrum in the AdS
case. For generic values of $\eta $ and $M$ the spectral function entering
the K\"all\'en-Lehmann representation for $\varphi$ is featureless. For
certain
values of $\eta $ however (for instance when $\eta$ is close to
$\eta_b(M)$), it develops a narrow and high peak, which corresponds
to excitations with sharply localized mass.

Concluding this preliminary part, we would like to fix the action
for the field $\Phi $. It reads
\beeq
S \, = \, \ha \int_{ \cal M}  \, \sqrt{G} dx^{(s+2)} \left[ G^{AB}
\de_A \Phi \, \de_B \Phi \, - \, M^2 \, \Phi^2 \right] \, - \,  \ha \int_{\de
{\cal M}} \sqrt{G_{\rm in}}
dx^{(s+1)} \, \eta \, \Phi^2 \, .
\label{act}
\end{equation}
The bulk space $\M$ has a Lorentzian metric $G_{AB}$; its time-like boundary
$\der \M$ (a $s$-brane) inherits the induced metric $G_{\rm in}$.
The variation of $S$ gives both the equation of motion
\beeq
G^{AB} \, \nabla_A \nabla_B \, \Phi \, + \, M^2 \, \Phi \, = \, 0\, ,
\label{eqm}
\end{equation}
and the boundary condition
\beeq
{ \left(n^A \nabla_A \, \Phi \, - \, \eta \, \Phi \right) }\vert_{\der
{\cal M}} \,
= \, 0 \, , \qquad  \eta \in \mathbb{R} \, ,
\label{bcg}
\end{equation}
where $n^A$ is the unit normal of $\der \M$.

The paper is organized as follows:
in the next section we take $G$ to be flat, whereas in section 3 we
consider the AdS metric.
The last section is devoted to our conclusions.

\section{\bf Flat bulk space}

The purpose of this section is to illustrate, in its simplest form,
a general mechanism for inducing quantum fields on a brane. Let us
consider the manifold $\M = \R^{s+1} \times \R_+$, where
$\R_+$ is the half line $\{y\in \R \, :\, y>0 \}$. We adopt the coordinates
$(x,y) \in \R^{s+1} \times \R_+$ and the notations
$x \equiv (x^0,x^1,...,x^s) = (x^0,\bx)$ and
$p \equiv (p^0,p^1,...,p^s) = (p^0,\bp)$. Our first task will be to
analyze the free scalar quantum field $\Phi (x,y)$, which
propagates on $\M = \R^{s+1} \times \R_+$ equipped with the flat metric
\beeq
G_{A B} = \begin{pmatrix} g & 0 \\ 0 & -1\end{pmatrix}\, ,
\qquad {\rm diag}\, g = (1,-1,...,-1) \, .
\label{met}
\end{equation}
The boundary $\der M$ coincides with the $s+1$-dimensional Minkowski space
$\mathbf{M}_{s+1} \equiv \{\R^{s+1},\, g\}$ and represents a $s$-brane.
Eq. (\ref{eqm}) gives rise to
\beeq
(\Box_G + M^2) \Phi (x,y) = 0 \, ,
\label{eqm1}
\end{equation}
where $\Box_G$ is the Laplacian associated with (\ref{met}) and
$M\geq 0$ is the mass. Let us introduce
\beeq
K \equiv -\sum_{i=1}^s \der_i^2 - \der_y^2 + M^2 \, .
\label{opK}
\end{equation}
Then
\beeq
\Box_G + M^2 = \der_0^2 + K\, ,
\end{equation}
and the quantization of Eq. (\ref{eqm1}) requires the study of
the operator $K$. In order to apply the standard and well known procedure,
one needs a self-adjoint positive $K$. The subtle point in analyzing
$K$ concerns the term $-\der_y^2$, which is defined
on the half line $\R_+$. We recall \cite{ReSi} in this respect that
K is not self-adjoint but only Hermitian
in the space $C_0^\infty (\R^s\times \R_+)$ of infinitely differentiable
functions
with compact support. The relative deficiency indices are however equal,
because $K$ commutes with the complex
conjugation. Therefore, $K$ admits self-adjoint extensions. All of them are
parametrized by
a real parameter $\eta$ and correspond to the boundary condition
\beeq
\lim_{y \downarrow 0} ( \der_y - \eta )\Phi (x,y) = 0 \, ,
\label{bc}
\end{equation}
which is just eq.(\ref{bcg}) applied to the case we are considering.
The parameter $\eta $ has dimension of mass. Eq.(\ref{bc}) represents the
so called
mixed boundary condition (in the limit $\eta \to 0$ and $\eta \to \infty $
one recovers the familiar Neumann and Dirichlet boundary conditions
respectively).
The associated self-adjoint operator will be denoted by $K_\eta$.

At this point,
the main step in constructing the quantum field $\Phi$
is to identify a complete orthonormal set of eigenfunctions of $K_\eta$. For
this purpose we first focus on the operator $-\der_y^2$ and consider
the family of functions
\beeq
\B = \begin{cases}
\{\psi (y,\lambda)  \, :\, y,\, \lambda \in \R_+ \} &
\text {if $\eta \geq 0$}\, , \\
\{\psi (y,\lambda) ,\, \psi_b (y) \, :\, y,\, \lambda \in \R_+ \} &
\text{if $\eta < 0$} \, ,
\end{cases}
\label{basis1}
\end{equation}
where
\beeq
\psi (y,\lambda)  = \e^{-i\lambda y} + B(\lambda )\e^{i\lambda y}\, , \qquad
B(\lambda ) = \frac{\lambda - i\eta}{\lambda + i\eta}\, ,
\label{ss}
\end{equation}
and
\beeq
\psi_b (y) = \sqrt {2|\eta|}\, \e^{\eta y} \, .
\label{bs}
\end{equation}
The elements of $\B$ satisfy the boundary condition
\beeq
\lim_{y \downarrow 0} ( \der_y - \eta )\psi (y) = 0 \, ,
\label{bc1}
\end{equation}
and have transparent physical interpretation in the
quantum mechanical problem defined by the Hamiltonian $-\der_y^2$ on $\R_+$:
the functions $\psi (y, \lambda)$ represent scattering
states ($B(\lambda )$ is the reflection coefficient
from the boundary at $y=0$), whereas $\psi_b$ describes the bound state
(with energy $- \eta^2 $), present for $\eta < 0$.
We will refer in what follows to $\psi_b$ as boundary state and we will show
that it gives an essential contribution to
the quantum field induced on the boundary $\mathbf{M}_{s+1}$.

One can verify the following orthogonality
and completeness relations:
\beeq
\int_0^\infty dy\, {\overline \psi}(y,\lambda_1)
\psi (y,\lambda_2) = 2\pi\, \delta (\lambda_1-\lambda_2) \, ,
\qquad \lambda_1,\, \lambda_2 \in \R_+ \, ,
\label{ort1}
\end{equation}
\beeq
\int_0^\infty dy\, {\overline \psi}(y,\lambda )
\psi_b (y) = 0 \, ,
\qquad \lambda \in \R_+ \, ,\quad \eta < 0\, ,
\label{ort2}
\end{equation}
\beeq
\int_0^\infty \frac{d\lambda}{2\pi} \,{\overline \psi}(y_1,\lambda )
{\psi}(y_2,\lambda) +
\theta (-\eta ) \psi_b(y_1) \psi_b(y_2) =
\delta (y_1-y_2) \, , \qquad y_1,\, y_2 \in \R_+ \, ,
\label{compl}
\end{equation}
where
\beeq
\theta (\alpha ) = \left\{ \begin{array}{cc} 1 & \mbox{if } \alpha > 0\, , \\
0 & \mbox{if } \alpha \leq 0 \, .
\end{array} \right.
\label{theta}
\end{equation}
Therefore, $\B$ is a complete orthonormal set of eigenfunctions
for the operator $-\der_y^2$. Accordingly,
\beeq
\B(K_\eta) = \begin{cases}
\{\e^{-i\bp \bx }\psi (y,\lambda)  \, :\, y,\, \lambda \in \R_+,\, \,
\bx,\, \bp \in \R^s \} &
\text {if $\eta \geq 0$}\, , \\
\{\e^{-i\bp \bx }\psi (y,\lambda) ,\, \e^{-i\bp \bx }\psi_b (y)
\, :\, y,\, \lambda \in \R_+,\, \,  \bx,\, \bp \in \R^s \} & \text{if $\eta
< 0$} \, ,
\end{cases}
\label{basis2}
\end{equation}
is such a set for the operator $K_\eta$. For the spectrum one has
\beeq
\Si(K_\eta) = \begin{cases}
\{\kappa \in \R\, :\, \kappa \geq M^2 \} &
\text {if $\eta \geq 0$}\, , \\
\{\kappa \in \R\, :\, \kappa \geq M^2 - \eta^2\} & \text{if $\eta < 0$}\, .
\end{cases}
\label{spec}
\end{equation}
Therefore, $K_\eta$ is positive for any $\eta \geq 0$. In the range
$\eta < 0$, the requirement of positivity implies
\beeq
-M\leq \eta < 0 \, .
\label{posit}
\end{equation}
In order to avoid imaginary energies for the field $\Phi $, in what follows we
impose (\ref{posit}). Notice that for $M=0$, this condition excludes
negative values of $\eta $. We shall see in the next section that this is
not the case for the AdS space, which is essential in the Randall-Sundrum
scenario.

Now, we are in position to construct the local quantum field $\Phi$. The
content
of $\B(K_\eta )$ suggest the presence of two building blocks:
\bea
\phi (x,y) = \int_{-\infty}^\infty \frac{d^sp}{(2\pi)^s} \int_0^\infty
\frac{d\lambda}{2\pi}
\frac{1}{\sqrt {2\omega_{M^2 + \lambda^2}(\bp )}}\cdot \qquad \qquad \qquad
\qquad \qquad \quad \nb \\
\left[a^*(\bp,\lambda )\e^{i\omega_{M^2 + \lambda^2}(\bp )x^0-i\bx \bp
}\psi (y,\lambda)  +
a(\bp,\lambda )\e^{-i\omega_{M^2 + \lambda^2}(\bp )x^0+i\bp \bx }{\overline
\psi}(y,\lambda)\right ] \, ,
\qquad -M\leq \eta \, ,
\label{f1}
\ena
and
\bea
\chi (x,y) = \int_{-\infty}^\infty \frac{d^sp}{(2\pi)^s}
\frac{1}{\sqrt {2\omega_{M^2 - \eta^2}(\bp )}}\cdot \qquad \qquad \qquad
\qquad \qquad \qquad \quad  \nb \\
\left[b^*(\bp)\e^{i\omega_{M^2 - \eta^2}(\bp )x^0-i\bx \bp }\psi_b (y) +
b(\bp)\e^{-i\omega_{M^2 - \eta^2}(\bp )x^0+i\bp \bx }{\overline \psi}_b
(y)\right ] \, ,
\qquad -M\leq \eta <0\, ,
\label{f2}
\ena
where
\beeq
\omega_{m^2}(\bp ) = \sqrt {\bp^2 + m^2} \, .
\label{disp}
\end{equation}
We assume also that
$\{a^*(\bp,\lambda ),\, a(\bp,\lambda )\}$ commute with  $\{b^*(\bp),\,
b(\bp)\}$
and satisfy the canonical commutation relations:
\bea
&&[a(\bp_1,\lambda_1 )\, ,\, a^*(\bp_2,\lambda_2 )] =
(2\pi)^{s+1}\delta (\bp_1-\bp_2)\delta (\lambda_1-\lambda_2) \, , \nb \\
&&[a(\bp_1,\lambda_1 )\, ,\, a(\bp_2,\lambda_2 )] =
[a^*(\bp_1,\lambda_1 )\, ,\, a^*(\bp_2,\lambda_2 )] = 0\, ,
\label{ccr1}
\ena
\bea
&&[b(\bp_1)\, ,\, b^*(\bp_2)] =
(2\pi)^s\delta (\bp_1-\bp_2) \, , \nb \\
&&[b(\bp_1)\, ,\, b(\bp_2)] = [b^*(\bp_1)\, ,\, b^*(\bp_2)] = 0\, .
\label{ccr2}
\ena
Let us summarize the basic properties of the fields (\ref{f1},\ref{f2}).
By construction both $\phi$ and $\chi$ satisfy Eqs. (\ref{eqm1}, \ref{bc}).
At this point, the requirement of local commutativity plays a crucial role.
Indeed, one can verify that:
\begin{itemize}
\item {(i)} $\phi $ is local for $\eta \geq 0$;
\item {(ii)} neither $\phi $ nor $\chi $ are local for $-M\leq \eta <0$,
however
their sum $\phi + \chi$ is local.
\end{itemize}
Therefore, the local field in $\R^{s+1}\times \R_+$ we are looking for, reads
\beeq
\Phi (x,y) = \begin{cases}
\phi (x,y) &
\text {if $\eta \geq 0$}\, , \\
\phi (x,y) + \chi (x,y) & \text{if $-M\leq \eta < 0$} \, .
\end{cases}
\label{f3}
\end{equation}
The completeness of the system $\B(K_\eta )$ implies that
that above defined $\Phi $ satisfies
also the canonical commutation relation
\beeq
\left [(\der_0\Phi )(x^0,\bx_1,y_1)\, ,\, \Phi (x^0,\bx_2,y_2)\right ]
= -i\delta (\bx_1-\bx_2)\delta(y_1-y_2) \, .
\label{ccr3}
\end{equation}

The two-point vacuum expectation values (Wightman functions) of the fields
$\phi $ and $\chi $ in the Fock representation of the algebra
(\ref{ccr1},\ref{ccr2}) are easily derived:
\beeq
\langle \phi (x_1,y_1) \phi(x_2,y_2)\rangle_0 =
\int_0^\infty \frac{d\lambda}{2\pi} {\overline \psi}(y_1,\lambda )
\psi (y_2,\lambda ) \, W_{M^2+\lambda^2}(x_{12}) \, ,
\label{cor1}
\end{equation}
and
\beeq
\langle \chi (x_1,y_1) \chi(x_2,y_2)\rangle_0 =
{\overline \psi}_b (y_1)
\psi_b (y_2) \, W_{M^2-\eta^2}(x_{12}) \, ,
\label{cor2}
\end{equation}
where $x_{12} \equiv x_1-x_2$ and
\beeq
W_{m^2}(x) = \int_{-\infty}^\infty \frac{d^sp}{(2\pi)^s} \,
\frac{1}{2\omega_{m^2}(\bp )} \e^{-i\omega_{m^2}(\bp )x^0 + i\bp \bx}
= \int_{-\infty}^\infty \frac{d^{(s+1)}p}{(2\pi)^{(s+1)}} \,
\e^{-ipx} \theta (p^0) 2\pi \delta (p^2 - m^2) \,
\label{sc}
\end{equation}
is the two-point scalar function of mass $m^2$ in $\mathbf{M}_{s+1}$. Combining
Eqs.(\ref{ss},\ref{bs},\ref{f3},\ref{cor1},\ref{cor2}), one gets
\bea
\langle \Phi (x_1,y_1) \Phi(x_2,y_2)\rangle_0 =
\qquad \qquad \qquad \qquad \qquad \qquad
\nb \\
\int_0^\infty \frac{d\lambda}{2\pi}\, 2
\left (\cos\lambda y_{12} +
\frac{\lambda^2-\eta^2}{\lambda^2+\eta^2}\cos\lambda {\widetilde y}_{12}
+ \frac{2\lambda \eta}{\lambda^2+\eta^2}\sin\lambda {\widetilde
y}_{12}\right )
W_{M^2+\lambda^2}(x_{12}) + \nb \\
2\theta (-\eta )|\eta|\e^{\eta {\widetilde y}_{12}}\,
W_{M^2-\eta^2}(x_{12}) \, ,
\qquad \qquad \qquad \qquad \qquad \quad
\label{cor3}
\ena
where ${\widetilde y}_{12} \equiv y_1+y_2$. Since $\Phi $ is free,
its $n$-point correlators are expressed in a standard way in terms of
(\ref{cor3}),
which completes the construction of the local bulk field $\Phi$, satisfying
Eqs. (\ref{eqm1},\ref{bc}).

So, we are left with the problem of determining the field $\varphi $
induced by $\Phi $
on the boundary $\bM_{s+1}$. For this purpose we consider
\beeq
\langle \varphi (x_1)\cdots \varphi (x_n)\rangle_0 =
\lim_{y_i \downarrow 0} \langle \Phi (x_1,y_1)\cdots \Phi(x_n,y_n)\rangle_0
\, ,
\label{cor4}
\end{equation}
It follows from Eq. (\ref{cor3}) that
$\langle \varphi (x_1)\cdots \varphi (x_n)\rangle_0$
are well defined correlators, which satisfy all
standard requirements (Poincar\'e covariance, local commutativity,
positivity and the
spectral condition) and therefore uniquely determine, via the
reconstruction theorem \cite{SW},
the induced field $\varphi$. The basic features of $\varphi $ are encoded
in the two-point function
\bea
G_{{}_{W}}(x_{12}) =
\langle \varphi (x_1)\varphi (x_2)\rangle_0 =
\lim_{y_i \downarrow 0} \langle \Phi (x_1,y_1)\Phi(x_2,y_2)\rangle_0 =
\qquad \nb \\
\int_0^\infty \frac{d\lambda}{2\pi}\, \frac{4\lambda^2}{\lambda^2+\eta^2}
W_{M^2+\lambda^2}(x_{12}) +
2\theta (-\eta )|\eta|\, W_{M^2-\eta^2}(x_{12}) \, .
\label{cor5}
\ena
Notice that $G_{{}_{W}}$ vanishes in the Dirichlet case ($|\eta |= \infty $),
as it should be. Changing variables, Eq. (\ref{cor5}) gives
the following K\"all\'en-Lehmann representation
\beeq
G_{{}_{W}}(x_{12}) \, = \,
\int_0^\infty d\lambda^2 \, \varrho (\lambda^2 )\, W_{\lambda^2}(x_{12}) +
2\theta (-\eta )|\eta|\, W_{M^2-\eta^2}(x_{12}) \, ,
\label{cor6}
\end{equation}
where
\beeq
\varrho (\lambda^2 ) =
\theta (\lambda^2 - M^2) \frac{\sqrt {\lambda^2 -
M^2}}{\pi(\lambda^2+\eta^2-M^2)} \, .
\label{klm}
\end{equation}
Obviously $\varphi $ does not satisfy the free Klein-Gordon equation.
Since $\varrho $ defines a polynomialy bounded positive measure on
$[0,+\infty )$, from Eq. (\ref{cor6}) we deduce that $\varphi $ is a
generalized free field \cite{SW}. We stress that in contrast
to the bulk field $\Phi $, the induced field $\varphi $ is not a canonical
one.
In this sense $\varphi $ cannot be derived from a local Lagrangian
on the brane.

The evaluation of the integral in Eq. (\ref{cor6}) is straightforward in
momentum space.
Performing a Fourier transformation, one finds
\beeq
{\widehat G}_{{}_{W}}(p) \, = \, 2\theta (p^0) \left [ \theta (p^2-M^2)
\frac{\sqrt {p^2-M^2}}{p^2-M^2 + \eta^2} + \theta (-\eta) |\eta | 2\pi
\delta(p^2 -M^2 + \eta^2) \right ] \, .
\label{ft1}
\end{equation}
Eq. (\ref{ft1}) provides a complete information about the mass spectrum of
$\varphi $. For $\eta\geq 0$ the spectrum is continuous and bounded from
below
by $M$, thus exhibiting a mass gap when $M>0$. Such a behavior is expected
on general grounds.
In the range $-M\leq \eta <0$ the mass spectrum has in addition a
discrete  point-like contribution, stemming from the boundary state
(\ref{bs}). The representation of the Poincar\'e group in this range
is a superposition of a continuum mass representation defined by $\varrho $
and a standard particle representation of mass $M^2-\eta^2$.
The particular case
\beeq
\eta = \eta_b \equiv -M \, ,
\label{etab1}
\end{equation}
deserves special attention.
In spite of the fact that only short range interactions
($M\not=0$) are
present in the bulk space $\R^{s+1}\times \R_+$, a long range interaction
($M^2-\eta^2 = 0$) is induced on its boundary $\mathbf{M}_{s+1}$. The
physical origin
of the remarkable behavior when $-M\leq \eta <0$ is a sort of attraction of
the bulk field by the boundary. We expect such boundary interactions to be
universal
and therefore essential in brane physics in general. In fact, we will show
in the next section that the above phenomena take place also in
AdS background, thus being relevant for the Randall-Sundrum
framework as well.

It is instructive to derive also the propagator of the induced field. From
Eq. (\ref{cor6}) one obtains
\beeq
G_{{}_{F}}(x_{12}) \, = \,i \langle T \varphi (x_1)\varphi (x_2)\rangle_0 =
\int_0^\infty d\lambda^2 \, \varrho (\lambda^2 )\, \Delta_{\lambda^2}(x_{12}) +
2\theta (-\eta )|\eta|\, \Delta_{M^2-\eta^2}(x_{12}) \, ,
\label{fey}
\end{equation}
where $T$ indicates time ordering and
\beeq
\Delta_{m^2}(x) \, = \, - \int_{-\infty}^\infty
\frac{d^{(s+1)}p}{(2\pi)^{(s+1)}} \,
\frac{\e^{-ipx}}{p^2 - m^2 + i\varepsilon }  \, .
\label{tsc}
\end{equation}
is the well known scalar propagator. The Fourier transform of $G_{{}_{F}}$
reads
\beeq
{\widehat G}_{{}_{F}}(p) \, = \, - \frac{\sqrt{M^2 - p^2} - \eta}{p^2 -M^2
+ \eta^2 + i \epsilon } \, ,
\label{fey1}
\end{equation}
and is convenient for perturbative calculations on the brane.

Let us determine finally the potential $V(r)$ between two static
sources on the brane at distance $r = |\bx |$. For $s=3$ one finds
\beeq
V(r) = \int_{-\infty}^{\infty}dx^0  G_{{}_{F}}(x^0, \bx) =
2\int_M^{\infty }d\lambda \, \lambda \varrho (\lambda^2 )\,
\frac{e^{-\lambda r}}{4\pi r}
+ 2\theta (-\eta ) |\eta | \frac{e^{-r\sqrt{M^2-\eta^2}}}{4\pi r} \, .
\label{pot1}
\end{equation}
Combining Eqs. (\ref{klm},\ref{pot1}), one easily gets the estimate
\beeq
V(r) \leq \sup_{\lambda \geq M}[\varrho (\lambda )]\, \frac{e^{-Mr}}{4\pi r^2}
+ 2\theta (-\eta ) |\eta | \frac{e^{-r\sqrt{M^2-\eta^2}}}{4\pi r} \, .
\label{pot2}
\end{equation}
One can further study the dependence of
$\sup_{\lambda \geq M}[\varrho (\lambda )]$ on $M$ and $\eta$.

This concludes our investigation of the flat bulk space, which provides useful
intuition for the AdS case.

\section{\bf Anti-de Sitter bulk space}

We keep below the bulk manifold $\M = \R^{s+1} \times \R_+$, but now
equipped with the anti-de Sitter (AdS) metric
\beeq
ds^2 \, = \, G_{A B} \, dx^A dx^B \, =
\, \e^{-2ay} \, g_{\mu \nu} \, dx^\mu dx^\nu \, - \, dy^2 \, ,
\qquad a>0 \, ,
\label{met1}
\end{equation}
with  a cosmological constant $\Lambda = - 6 a^2$.
The boundary $\der \M$ still coincides with $\mathbf{M}_{s+1}$ and
the problem is to construct and investigate the scalar quantum field $\Phi $,
satisfying Eq. (\ref{eqm1}) with the metric (\ref{met1}) and the boundary
condition (\ref{bc}). Following the
procedure developed in the previous section, we have to study the operator
\beeq
K = -\sum_{i=1}^s \der_i^2 - \e^{(s-1)ay}\der_y \e^{-(s+1)ay}\der_y +
\e^{-2ay}M^2 \, .
\label{opK2}
\end{equation}
As in the flat case, the main point is to solve
\beeq
\left [- \e^{(s-1)ay}\der_y \e^{-(s+1)ay}\der_y + \e^{-2ay}M^2 \right
]\psi(y,\lambda ) =
\lambda^2 \psi(y,\lambda )\, ,
\label{spp}
\end{equation}
with the boundary condition (\ref{bc1}). Eqs. (\ref{spp},\ref{bc1})
give rise to a well studied \cite{T}, \cite{N} singular boundary value
problem,
related to Bessel's equation.
It is worth stressing that there is no freedom to impose
any boundary condition at $y=\infty$, which is the singular point of the
problem. Setting
\beeq
\nu = \sqrt {\frac{(s+1)^2}{4} + \frac{M^2}{a^2}} \, ,  \qquad
\eta_b = \frac{(s+1-2\nu)}{2}a \, ,
\label{ni}
\end{equation}
the complete orthonormal set of eigenfunctions for $s\geq 3$ reads
\beeq
\B = \begin{cases}
\{\psi (y,\lambda) \, :\, y,\, \lambda \in \R_+ \} &
\text {if $\eta \not= \eta_b$}\, , \\
\{\psi (y,\lambda),\, \psi_b (y) \, :\, y,\, \lambda \in \R_+ \} &
\text{if $\eta = \eta_b$} \,
\end{cases}
\label{basisd1}
\end{equation}
where
\beeq
\psi (y,\lambda) = \e^{\frac{(s+1)}{2}ay}
\left [J_\nu (\lambda a^{-1} \e^{ay})
{\widetilde Y}_\nu (\lambda a^{-1}) -
Y_\nu(\lambda a^{-1} \e^{ay})
{\widetilde J}_\nu (\lambda a^{-1})\right ] \, ,
\label{bess}
\end{equation}
\beeq
{\widetilde Z}_\nu (\zeta ) = \frac{1}{2\sqrt{1+\widetilde{\eta}^2}}
\left[(1-2\widetilde {\eta})Z_\nu(\zeta ) + 2\zeta Z_\nu^\prime (\zeta )
\right ] \, ,
\qquad \widetilde {\eta} = \frac{\eta}{a} - \frac{s}{2} \, ,
\label{til}
\end{equation}
$Z_\nu$ denoting the Bessel function $J_\nu$ or $Y_\nu$ of first and
second kind respectively. Finally, the boundary state
\beeq
\psi_b(y) = \sqrt {2a(\nu -1)}\, \e^{\eta_by}
\label{bs1}
\end{equation}
is in this case a zero mode of Eq. (\ref{spp}). The completeness relation
of the
system (\ref{basisd1}) is expressed by
\beeq
\int_0^\infty d\lambda\, \sigma (\lambda )
{\overline \psi} (y_1,\lambda) \psi (y_2,\lambda) +
\delta_{\eta \eta_b}{\overline \psi}_b(y_1)\, \psi_b(y_2) =
\frac{1}{\mu (y_1)}\, \delta (y_{12}) \, ,
\label{compl1}
\end{equation}
where $\mu $ and $\sigma $ are the following measures
\beeq
\mu (y) = \e^{(1-s)ay}\, , \qquad
\sigma (\lambda ) =
\frac{\lambda a^{-1}}{{\widetilde J}_\nu^2 (\lambda a^{-1}) +
{\widetilde Y}_\nu^2 (\lambda a^{-1})} \, ,
\label{mesures}
\end{equation}
on $[0,\infty)$.
Moreover, one has also the orthogonality relations
\beeq
\int_0^\infty dy\, \mu (y)\, {\overline \psi} (y,\lambda_1) \psi
(y,\lambda_2) =
\frac{1}{\sigma (\lambda_1)}\delta(\lambda_{12}) \, ,
\qquad \lambda_1,\, \lambda_2 \in \R_+ \, ,
\label{ort3}
\end{equation}
\beeq
\int_0^\infty dy\, \mu(y)\, {\overline \psi} (y,\lambda )\psi_b(y) = 0 \, ,
\qquad
\lambda \in \R_+ \, .
\label{ort4}
\end{equation}
The above analysis implies that
\beeq
\B(K_\eta) = \begin{cases}
\{\e^{-i\bp \bx }\psi (y,\lambda )  \, :\, y,\, \lambda \in \R_+,\, \,
\bx,\, \bp \in \R^s \} &
\text {if $\eta \not= \eta_b$}\, , \\
\{\e^{-i\bp \bx }\psi (y,\lambda ) ,\, \e^{-i\bp \bx }\psi_b (y)
\, :\, y,\, \lambda \in \R_+,\, \,  \bx,\, \bp \in \R^s \} &
\text{if $\eta = \eta_b$} \, ,
\end{cases}
\label{basis4}
\end{equation}
is a complete orthonormal set for the self-adjoint extension $K_\eta $
of the operator (\ref{opK2}). For the spectrum one has
\beeq
\Si(K_\eta) = \begin{cases}
\{\kappa \in \R\, :\, \kappa > 0 \} &
\text {if $\eta \not= \eta_b$}\, , \\
\{\kappa \in \R\, :\, \kappa \geq 0\} & \text{if $\eta = \eta_b$}\, ,
\end{cases}
\label{specd}
\end{equation}
showing that $K_\eta$ is positive for any $\eta \in \R$. Comparing to the
flat case,
we see that the AdS background has an interesting feature: the energy
spectrum of the field $\Phi $ is real and positive for any $\eta \in \R$.

It follows from Eq. (\ref{basis4}) that $\Phi $ is given by the superposition
\beeq
\Phi (x,y) = \phi (x,y) + \delta_{\eta \eta_b} \chi (x,y) \, ,
\label{ads}
\end{equation}
where
\bea
\phi (x,y) = \int_{-\infty}^\infty \frac{d^sp}{(2\pi)^s} \int_0^\infty
d\lambda \sigma (\lambda )
\frac{1}{\sqrt {2\omega_{\lambda^2}(\bp )}}\cdot \qquad \qquad  \nb \\
\left [a^*(\bp,\lambda )\e^{i\omega_{\lambda^2}(\bp )x^0-i\bx \bp }\psi
(y,\lambda)  +
a(\bp,\lambda )\e^{-i\omega_{\lambda^2}(\bp )x^0+i\bp \bx }{\overline
\psi}(y,\lambda)\right ] \, ,
\label{f3d}
\ena
\bea
\chi (x,y) = \int_{-\infty}^\infty \frac{d^sp}{(2\pi)^s}
\frac{1}{\sqrt {2\omega_0(\bp )}}\cdot
\qquad \qquad \nb \\
\left[b^*(\bp)\e^{i\omega_0(\bp )x^0-i\bx \bp }\psi_b (y) +
b(\bp)\e^{-i\omega_0(\bp )x^0+i\bp \bx }{\overline \psi}_b (y)\right ] \, .
\label{f4}
\ena
Here $\{a^*(\bp,\lambda ),\, a(\bp,\lambda )\}$ and $\{b^*(\bp),\,
b(\bp)\}$ satisfy
the commutation relations (\ref{ccr1},\ref{ccr2}) with the substitution
\beeq
2\pi \delta (\lambda_1-\lambda_2) \longmapsto
\frac{1}{\sigma (\lambda_1)}\, \delta (\lambda_1-\lambda_2) \,
\end{equation}
in (\ref{ccr1}). Using Eq. (\ref{compl1}) one easily verifies that $\Phi $
satisfies
Eq. (\ref{ccr3}), being therefore a canonical field. Its two point vacuum
expectation
value admits the representation
\bea
\langle \Phi (x_1,y_1) \Phi(x_2,y_2)\rangle_0 =
\qquad \qquad \qquad \qquad \qquad
\nb \\
\int_0^\infty d\lambda \, \sigma (\lambda )
{\overline \psi}(y_1,\lambda )
\psi (y_2,\lambda ) \, W_{\lambda^2}(x_{12}) +
\delta_{\eta\eta_b} {\overline \psi}_b (y_1)
\psi_b (y_2) \, W_0(x_{12}) \, .
\qquad
\label{cor7}
\ena
{}From Eq. (\ref{cor7}) one gets
\bea
G_{{}_W}(x_{12}) =
\langle \varphi (x_1)\varphi (x_2)\rangle_0 =
\lim_{y_i \downarrow 0} \langle \Phi (x_1,y_1)\Phi(x_2,y_2)\rangle_0 = \qquad
\qquad \nb \\
\int_0^\infty d\lambda \, \sigma (\lambda )\, {\overline \psi}(0,\lambda )
\psi (0,\lambda )
W_{\lambda^2}(x_{12}) +
\delta_{\eta\eta_b} 2a(\nu -1)\, W_0(x_{12}) \, ,
\label{cor8}
\ena
which is the K\"all\'en-Lehmann representation for the induced field
$\varphi $ in the AdS case. It follows from Eqs. (\ref{bess},\ref{til}) that
$G_{{}_W}$ vanishes for $|\eta|=\infty$ (Dirichlet boundary condition). We
therefore
consider from now on only finite $\eta $. Using some
simple identities among Bessel functions, one finds
\beeq
G_{{}_W}(x_{12}) =
\int_0^\infty d\lambda \, \tau (\lambda )\, W_{\lambda^2}(x_{12}) +
\delta_{\eta\eta_b} 2a(\nu -1)\, W_0(x_{12}) \, ,
\label{cor9}
\end{equation}
with
\beeq
\tau (\lambda ) = \frac{16}{\pi^2}\,
\frac{\lambda a^{-1}}
{[(1-2{\widetilde \eta})J_\nu (\lambda a^{-1}) + 2\lambda a^{-1}
J_\nu^\prime (\lambda a^{-1})]^2 +
[(1-2{\widetilde \eta})Y_\nu (\lambda a^{-1}) + 2\lambda a^{-1}
Y_\nu^\prime (\lambda a^{-1})]^2} \, .
\label{tau}
\end{equation}
The $\lambda$-integral in Eq. (\ref{cor9}) is easily performed in
momentum space, giving the following result
\beeq
{\widehat G}_{{}_W}(p) = 2\pi \theta(p^0)
\left [ \frac{\theta(p^2)}{2 \sqrt{p^2}} \, \tau (\sqrt {p^2}) +
\delta_{\eta\eta_b} 2a(\nu -1)\, \delta (p^2) \right ] \, .
\label{cor10}
\end{equation}
Like in the flat space, in spite of the fact that $\Phi $ is a
canonical field, $\varphi $ is not. There is
no mass gap in the spectrum, which contains a zero-mass point-like
contribution only for the specific boundary condition $\eta = \eta_b$,
when the boundary state (\ref{bs1}) belongs to $\B(K_\eta )$.

\vskip 0.5  truecm
\begin{figure}[h]
\begin{picture}(13,13)
\put(100,110){$ \tau$}
\put(190,20){$\lambda /a$}
\put(360,60){$M/a$}
\end{picture}
\centerline{\epsfig{file=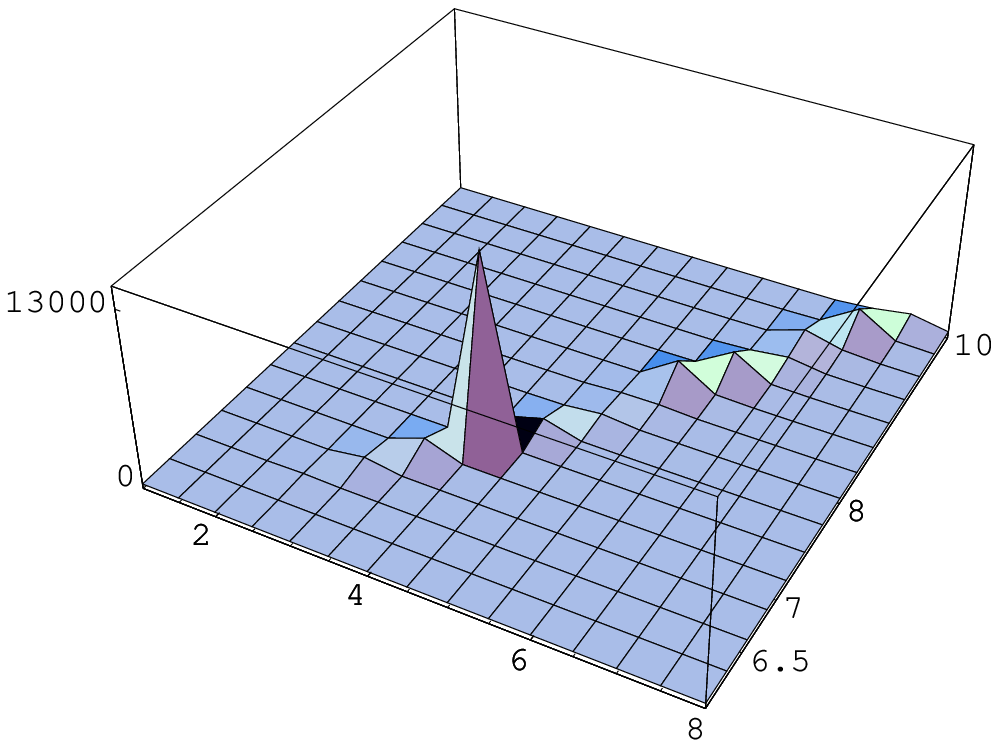,height=8cm,width=10cm}}
\centerline{ {\bf Figure 1.} The spectral function $\tau$ for $\eta =
\eta_b(6a)$.}
\end{figure}
\vskip 0.5 truecm

\vskip 0.5  truecm
\begin{figure}[h]
\begin{picture}(13,13)
\put(100,110){$ \tau$}
\put(190,20){$\lambda /a$}
\put(360,60){$\eta /a$}
\end{picture}
\centerline{\epsfig{file=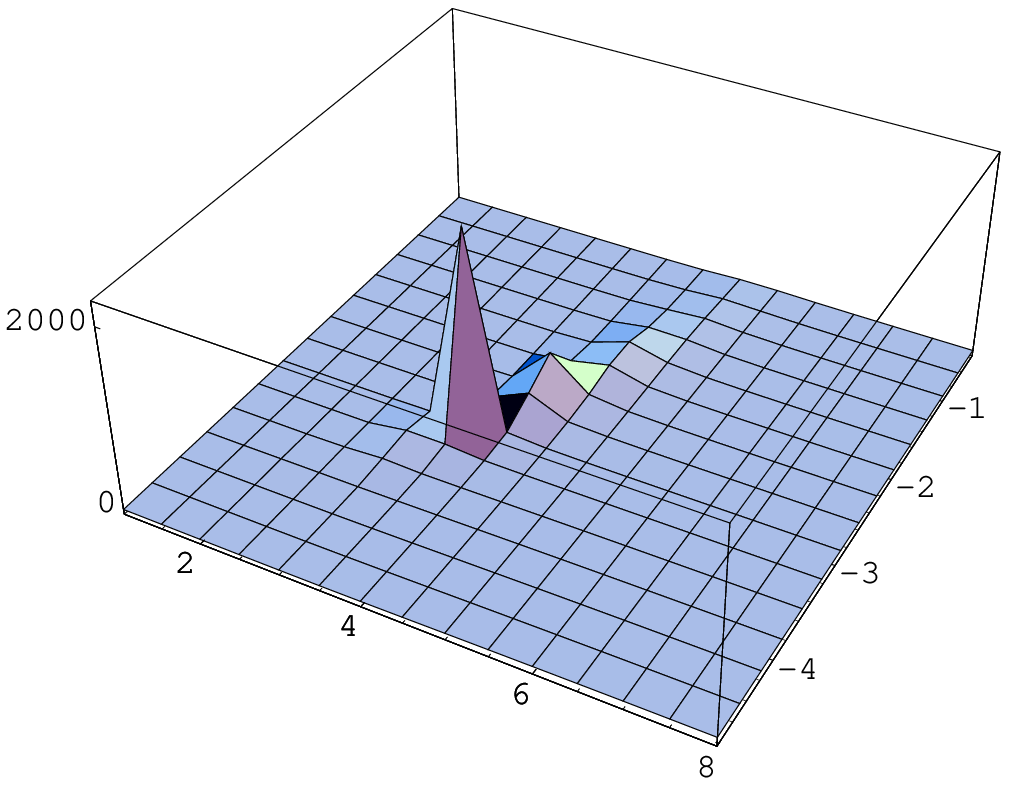,height=8cm,width=10cm}}
\centerline{ {\bf Figure 2.} The spectral function $\tau$ for $M = 6a$.}
\end{figure}
\vskip 0.5 truecm

It is hard to study analytically the behavior of the measure $\tau $
as a function of $\lambda $, $M$ and $\eta $. The numerical analysis
summarized in Figures 1 and 2 reveals a very interesting structure. For
generic
values of $M$ and $\eta$ the spectral function has monotonic behavior.
However,
when these parameters are close to the curve $\eta = \eta_b(M)$ in the
plane $(\eta, M)$,
a high peak appears in $\tau $, the rest of the spectrum being depressed.
As a results a partial
localization of a massive mode occurs on the brane. Some doubts have been
raised \cite{rub1}
about the stability of the corresponding excitations. In order to clarify the
issue, it is necessary to consider the propagator
\beeq
G_{{}_F}(x_{12}) = i\langle T\varphi (x_1) \varphi(x_2)\rangle_0 =
\int_0^\infty d\lambda \, \tau (\lambda )\, \Delta_{\lambda^2}(x_{12}) +
\delta_{\eta\eta_b} 2a(\nu -1)\, \Delta_0(x_{12}) \, . \qquad
\label{tcor}
\end{equation}
If our understanding of \cite{rub1} is correct, it is claimed
there that the Fourier transform
\beeq
{\widehat G}_{{}_F}(p) =
-\int_0^\infty d\lambda \, \tau (\lambda )\, \frac{1}{p^2-\lambda^2 +
i\epsilon} -
\delta_{\eta\eta_b} 2a(\nu -1)\, \frac{1}{p^2 + i\epsilon}
\label{tcorp}
\end{equation}
develops complex poles in the variable $p^2$. Differently from the flat
case (see Eq. (\ref{fey1})), the explicit form of
${\widehat G}_{{}_F}$ for generic $M$ and $\eta $ is quite complicated.
Nevertheless, using that $\tau $ is continuous and bounded on $\R_+$ for
any finite $\eta$,
we conclude that the $\lambda$-integral in Eq. (\ref{tcorp}) converges
when $p^2$ has a non-vanishing imaginary part even for $\epsilon = 0$.
This excludes the presence of complex poles in ${\widehat G}_{{}_F}$
on the physical sheet of the complex $p^2$-plane. Moreover, being
a generalized free field, $\varphi $ admits \cite{SW} a Fock representation
$\cal F$. The subspaces of $\cal F$ with different particle number are
orthogonal and invariant under the Hamiltonian. These features,
combined with translation invariance in
$\mathbf{M}_{s+1}$, forbid any decay process of $\varphi $.
In this sense $\varphi $ is a stable quantum field on the brane.

For $s=3$ the potential $V(r)$ between two static sources on the brane
at a distance $r = |\bx |$ is
\beeq
V(r) = \int_{-\infty}^{\infty}dx^0  G_{{}_{F}}(x^0, \bx) =
\int_0^{\infty }d\lambda \, \tau (\lambda )\, \frac{e^{-\lambda r}}{4\pi r}
+ \delta_{\eta\eta_b} 2a(\nu -1) \frac{1}{4\pi r} \, .
\label{pot3}
\end{equation}
Since $\tau $ is bounded, one has the estimate
\beeq
V(r) \leq \sup_{\lambda \geq 0}[\tau (\lambda )]\, \frac{1}{4\pi r^2}
+ \delta_{\eta\eta_b} 2a(\nu -1) \frac{1}{4\pi r} \, .
\label{pot4}
\end{equation}
Eq. (\ref{pot4}) holds for any $r>0$; for limited regions on
$\R_+$ and selected values of $M$ and $\eta $ one can derive
sharper estimates. Suppose for instance we are in the regime when
a high peak is present in the spectral function $\tau $. Then, even if at
very large
($r >> a^{-1}$) and very small ($r << a^{-1}$) distances $V(r)$ has a
power-like behavior,
there is an intermediate region in which $V(r)$ is well approximated by an
Yukawa type potential
$\sim \exp(-m r)/r$. The value of $m$ is given by the location of the peak
in $\tau$,
whereas the limits of validity of the Yukawa approximation are determined
by its width.

The cases $s=1,2$ and/or $-[a(s+1)/2]^2 \leq M^2 < 0$ can be investigated
in analogous way.

\section{Outlook and conclusions}

We studied a quantum scalar field of mass $M$ which propagates in a
manifold with boundary representing
a $s$-brane. The quantization of this model boils down to the solution of a
singular Sturm-Liouville
problem with a prescribed boundary condition on the brane, fixed in terms
of the parameter $\eta$.
Once the bulk 2-point function
$\langle \Phi (x_1,y_1) \Phi(x_2,y_2)\rangle_0$ has been computed,
the notion of induced quantum field $\varphi $ on the brane appears naturally,
using the limit of $y_i \to 0$. The resulting 2-point function uniquely
determines
$\varphi$, which turns out to be a generalized free field. Being
non-canonical, the dynamics of
$\varphi $ cannot be derived from a local Lagrangian defined on the brane.
The basic properties of the induced field are captured by the spectral
measure of the
corresponding K\"all\'en-Lehmann representation.
The importance of exploring the whole allowed range of $M$ and $\eta$ emerges
from our study: both in the flat and in the AdS case, the existence of a
boundary state strongly depends
on the values of these parameters. We have shown that under certain
conditions, short range bulk
interactions can induce long range forces on the brane. This phenomenon is
the consequence of
peculiar boundary interactions and to our knowledge has not been observed
before.
We have seen also that the spectral function $\tau (\lambda )$ in the AdS case
has quite complicated behavior. For certain values of $\eta $, it develops
a high
peak corresponding to excitations with sharply localized mass. This
feature
is very interesting and deserves further investigation.

Although we have focused in the present paper on a scalar field, our
approach applies with slight modifications to fields with higher spin as
well. For
instance, in the
gravitational case the linearized Einstein
equations lead to an operator similar to $K$ (see Eq. (\ref{opK2})) with
$M=0$.
In absence of matter on the brane, Israel's
junction conditions impose $\eta =0$. Observing that
$\eta_b(0) = 0$ (see Eq. (\ref{ni})), the boundary state is
present in the spectrum and is responsible for recovering standard
gravity on the brane \cite{RS2}. An interesting problem is the extension
of our analysis to the model described in \cite{grs}. The
boundary state there is not normalizable.
Nevertheless, at intermediate scales
the induced field reproduces four dimensional gravity.
This feature resembles the observed structure in the AdS
spectral function $\tau $ for $\eta$ close to $\eta_b(M)$.

Finally, is is natural to ask what happens when interactions in the bulk
and/or on
the brane are turned on. It will be interesting in this respect to perform
some
perturbative calculation based on the generalized free propagator
(\ref{tcor}).
We are currently working on this issue.

\bigskip
\medskip
{\bf Acknowledgments}

\bigskip
The collaboration of A. Liguori in deriving some of the results of section
2 is kindly
acknowledged. We also thank R. Rattazzi for his interest in this work and
for useful
discussions.

\end{document}